\documentclass[journal=jacsat,manuscript=article]{achemso}
\setkeys{acs}{articletitle = true}
\usepackage{amsmath}
\usepackage{mathrsfs}
\usepackage{natbib}
\usepackage{multicol}

\usepackage[version=3]{mhchem} % Formula subscripts using \ce{}
\usepackage[T1]{fontenc}       % Use modern font encodings
\author{Jeremy Flannery}
\affiliation{Institute for Quantum Computing, University of Waterloo, Ontario, Canada}
\alsoaffiliation{Physics and Astronomy, University of Waterloo, Ontario, Canada}
\email{jbflanne@uwaterloo.ca}

\author{Rubayet Al Maruf}
\affiliation{Institute for Quantum Computing, University of Waterloo, Ontario, Canada}
\alsoaffiliation{Electrical and Computer Engineering, University of Waterloo, Ontario, Canada}

\author{Taehyun Yoon}
\affiliation{Institute for Quantum Computing, University of Waterloo, Ontario, Canada}
\alsoaffiliation{Electrical and Computer Engineering, University of Waterloo, Ontario, Canada}

\author{Michal Bajcsy}
\affiliation{Institute for Quantum Computing, University of Waterloo, Ontario, Canada}
\alsoaffiliation{Physics and Astronomy, University of Waterloo, Ontario, Canada}
\alsoaffiliation{Electrical and Computer Engineering, University of Waterloo, Ontario, Canada}
\email{mbajcsy@uwaterloo.ca}

\title{Fabry-P\'{e}rot cavity formed with dielectric metasurfaces in a hollow-core fiber}

\abbreviations{}

\begin{document}

%\begin{tocentry}
%
%\includegraphics[]{TOCGraphic.pdf}
%
%%Some journals require a graphical entry for the Table of Contents.
%%This should be laid out ``print ready'' so that the sizing of the
%%text is correct.
%%
%%Inside the \texttt{tocentry} environment, the font used is Helvetica
%%8\,pt, as required by \emph{Journal of the American Chemical
%%Society}.
%%
%%The surrounding frame is 9\,cm by 3.5\,cm, which is the maximum
%%permitted for  \emph{Journal of the American Chemical Society}
%%graphical table of content entries. The box will not resize if the
%%content is too big: instead it will overflow the edge of the box.
%%
%%This box and the associated title will always be printed on a
%%separate page at the end of the document.
%%
%\end{tocentry}

\begin{abstract}
  We demonstrate a fiber-integrated Fabry-P\'{e}rot cavity formed by attaching a pair of dielectric metasurfaces to the ends of a hollow-core photonic-crystal fiber segment. The metasurfaces consist of perforated membranes designed as photonic-crystal slabs that act as planar mirrors but can potentially allow injection of gases through their holes into the hollow core of the fiber. We have so far observed cavities with finesse of $\sim11$ and Q factors  of $\sim4.5\times10^{5}$ but much higher values should be achievable with improved fabrication procedures. We expect this device to enable development of new fiber lasers, enhanced gas spectroscopy, and studies of fundamental light-matter interactions and non-linear optics.
  
\end{abstract}

Several notable demonstrations of strong light-matter interactions and nonlinear optical processes at low light levels in hollow-core fibers have been reported in recent years. They include all-optical switching with few-hundred photons \cite{Bajcsy2009} and stationary light pulses \cite{Blatt2016} in hollow-core photonic crystal fibers (HCPCFs) loaded with laser-cooled atoms, as well as cross-phase modulation with few photons \cite{Venkataraman2013} and single-photon broadband quantum memory \cite{Sprague2014} in HCPCFs filled with room-temperature alkali atoms. At the same time, there is potential to further enhance such processes by integrating a cavity into the hollow-core fiber, which would broaden the new horizons already opened by the hollow-core fiber platform.

In solid-core fibers, cavity integration can be realized with fiber Bragg gratings \cite{Barmenkov2006} implemented by periodically modulating the refractive index of the fiber material, and in particular, of the core. However, since the core of a HCPCF is empty and the fiber structure is designed to minimize the overlap between the guided mode and the glass forming the photonic crystal, integrating a quarter-wave Bragg mirror  into a HCPCF \cite{Flannery2017} has not yet been demonstrated experimentally. 

High-finesse cavities have also been reported with reflective coatings deposited on the ends of a solid-core fiber piece \cite{Obrzud2016}. Additionally, cavities in solid-core photonic-crystal fibers were earlier realized by pressing mirrors against the cleaved ends of a fiber section \cite{Hendrickson2007}. Unfortunately, sealing the face of a HCPCF with a multi-layer reflective coating or with a mirror would make it impossible to introduce gases into the fiber core after the cavity has been formed.

Metallic and dielectric metasurfaces, formed by large two-dimensional arrays of nano-scale patterns, have been extensively explored in the past decade due to their capabilities to manipulate light in previously unimaginable ways \cite{Yu2014, Arbabi2015, Mueller2017}. Metasurfaces realized by perforating dielectric membranes, also known as photonic crystal slabs \cite{Fan2002} or high-contrast gratings \cite{Chang-review2012}, offer the additional advantage of being permeable by gases and liquids \cite{Kilic2011} and thus make an almost obvious choice as a technology for forming a cavity inside a HCPCF.

Here, we report the realization of a Fabry-P\'{e}rot cavity integrated into a hollow core photonic crystal fiber using an approach relying on photonic crystal slabs \cite{Fan2002} that we recently proposed for integrating cavities into hollow-core anti-resonant reflection waveguides (ARROWs) \cite{Bappi2017}. In this cavity, schematically shown in Fig. 1a, the HCPCF provides tight transverse confinement of light to its hollow core, while the photonic crystal slabs acting as dielectric metasurface mirrors are mounted onto the ends of a fiber segment and provide longitudinal confinement of light along the axis of the fiber.

The photonic-crystal membranes used here are designed and fabricated in silicon nitride. This material choice was determined by the target wavelength of 852nm for our cavities, as we plan to eventually use the cavities for experiments with laser-cooled cesium atoms. The design can be adjusted for other wavelengths or dielectric materials.We grow the silicon nitride on a silicon wafer using low-pressure chemical vapor deposition (LPCVD) producing a film with refractive index of 2.26 at 852 nm. The reflection and transmission spectra of the membrane mirror are determined by the membrane thickness and refractive index, radius of the holes, spacing between the holes (lattice constant), and the shape of the holes. The potentially high reflectivity of these patterned membranes arises from the presence of internally guided resonance modes that are confined to the membrane by total internal reflection \cite{Fan2002}. Light trapped in these resonances can leak out by coupling
% The periodic index modulation provided by the holes in the membrane allows for coupling 
to external radiation modes.
% by means of phase matching. 
%Interference \cite{Fan2003} 
At particular resonant optical frequencies, the leaked light will interfere constructively in the backward direction and destructively in the forward direction with respect to the incident light  \cite{Fan2003}, which can, under ideal conditions, result in a structure acting as a perfect mirror.  

%destructive and constructive interference in the forward and backward incident light directions, respectively, resulting in complete reflection of the incident light.

\begin{figure}[htb!]
\centering
%\begin{subfigure}[b]{0.6\linewidth} %0.65
%\includegraphics[width=\linewidth]{Fibercavity.pdf}
%\caption{}\label{fig:fibercavity}
%\end{subfigure}
%\begin{subfigure}[b]{0.49\linewidth} %0.49
%\includegraphics[width=\linewidth]{PWvGaussSIM.pdf}
%\caption{}\label{fig:PWvsGauss}
%\end{subfigure}
%\begin{subfigure}[b]{0.49\linewidth} %0.49
%\includegraphics[width=\linewidth]{PWvGaussOptSIM.pdf}
%\caption{}\label{fig:PWvGaussOptSIM}
%\end{subfigure}
\includegraphics[width=\linewidth]{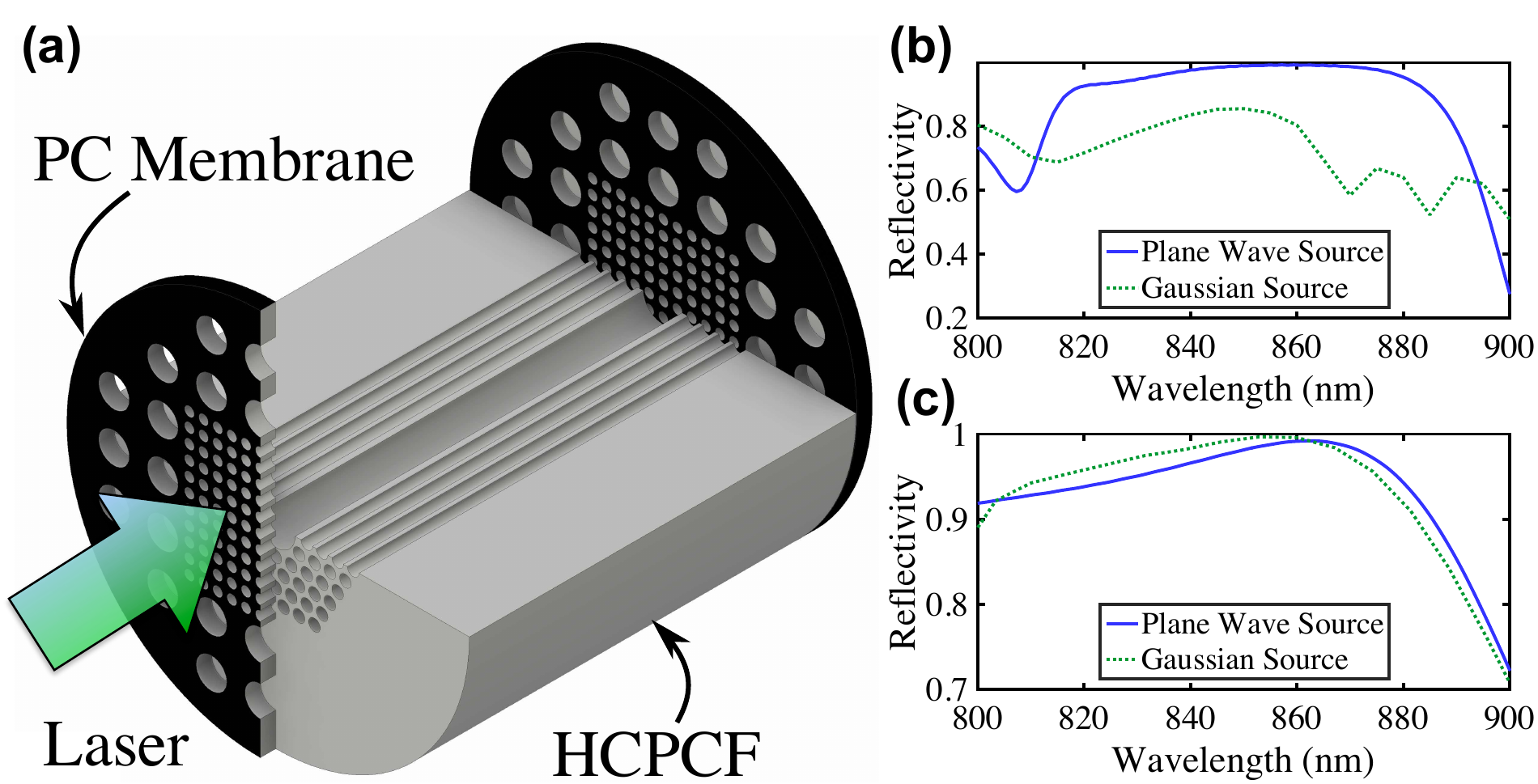}

\caption{(a) The HCPCF-integrated Fabry-P\'{e}rot cavity. The fiber serves as the cavity medium and confines light in the transverse direction, while a pair of dielectric metasurfaces (PC membranes) mounted on the ends of the fiber segment confine light longitudinally. (b) FDTD simulation results of the reflectivity produced for light incident perpendicularly on a PC membrane with lattice constant $a$=819 nm, hole radius $r$=347 nm, and film thickness $d$=500 nm. A plane wave and a Gaussian beam (with 2.75 $\mu$m waist radius) experience significantly different reflectivities. (c) The simulated reflectivity spectra for plane wave and Gaussian beam with membrane parameters chosen to optimize the reflectivity for the 2.75 $\mu$m-waist Gaussian source ($a$=680 nm, $r$=297 nm, $d$=369 nm).}
\end{figure}

%The main purpose behind such a fiber-integrated cavity, depicted in Fig. \ref{fig:fibercavity}, is that the presence of atomic vapor in the optical resonators may produce optical non-linearities resulting in enhanced light-matter interactions in cavity quantum electrodynamic (QED) experiments \cite{Bajcsy2009}. 
%While cold atoms have previously been loaded into the core of HCPCFs \cite{Christensen2008, Bajcsy2011}, the introduction of an atomic ensemble into the fiber hollow core of our fiber-cavities may be possible by use of the PC holes to pass through the dielectric metasurfaces. The proceeding section of this paper will outline the design, fabrication and assembly of these fiber-cavities. The later sections will describe the device performance and a means for tuning the cavity resonances.
%Our fiber-cavities are designed for the introduction of Cesium atoms, with a transition wavelength at 852 nm, and thus our PC mirrors are designed to be highly reflective at this wavelength. The material chosen for the PC membranes was silicon nitride, in order to minimize absorption in this near-IR region. Silicon nitride is grown on a silicon wafer using low-pressure chemical vapor deposition (LPCVD) producing a film with index of 2.26 at 852 nm.
In order to find the combination of parameters that will produce broad-band, near-unity reflectivity  
%a relatively broad, highly reflective region 
at the desired wavelength, we employ finite difference time domain (FDTD) simulations using a commercial software package (Lumerical) to determine the spectrum of the photonic-crystal mirrors. While high reflectivity for a broad range of wavelengths may not be a necessary requirement for some applications, it should make it easier to maintain high reflectivity at the target wavelength if imperfections arising in the fabrication process end up shifting the structure's central wavelength.
The photonic-crystal pattern used here is a simple square lattice of circular holes, in which three parameters dictate the full optical properties of the mirrors: lattice constant (hole pitch), hole radius, and thickness of the silicon nitride film. In previous studies \cite{Lousse2004}, highly reflective broad band regions have been found using computationally more efficient techniques, such as transfer matrix methods, however these methods assume the light incident on the structure is a plane wave. 

Here, on the other hand, the light incident on the membrane will be the fundamental mode of the HCPCF, which has a nearly Gaussian profile with a waist size similar to the radius of the hollow core. In this work, we used HC-800-02 fiber available from NKT Photonics, which has a mode field diameter of $\sim 5.5 \mu$m.
%The reflectivity of the photonic-crystal membrane can be highly dependent on the type of incident source. 
Fig. 1b shows the difference between the spectrum of a PC membrane assuming a plane wave source and a Gaussian source as the incident light. This illustrates that the reflective properties of the photonic-crystal membrane can be highly dependent on the transverse profile of the incident light, especially for the relatively tightly-confined fiber-guided modes.

As a result, our FDTD simulations have to simulate the whole mirror structure, as the translational periodicity and symmetries, which can be exploited to simplify the calculations for plane waves, are not available.   
This resulted in much larger computational requirements and the Particle Swarm Optimization (PSO) algorithm provided in the Lumerical package was used to search the parameter space in order to maximize reflectivity. A combination of parameters predicting a reflectivity $>99.9\%$ at 852nm was found to consist of a 680 nm lattice constant, hole radius of 297 nm, and membrane thickness of 369 nm (Fig. 1c), in which both a Gaussian source as well as a plane wave source resulted in relatively high reflectivities. It was also found that the reflectivity was $>99\%$ for Gaussian mode sizes with a waist radius $>2 \mu$m. However, given the multiple dimensions of the parameter space, other parameter combinations are likely to result in high reflectivities as well.

We fabricate the photonic-crystal membranes using electron-beam lithography and dry etching. The fabrication procedure begins by depositing a 40 nm thick aluminium hard mask onto the silicon nitride film that was grown on a silicon wafer. We then spin coat a 450 nm thick layer of ZEP520A (Zeon Chemicals) onto the chip. The hole pattern is written in the ZEP resist using e-beam lithography at 100 keV and developed using amyl-acetate. Reactive Ion Etching (RIE) is used to transfer the pattern onto the aluminum mask using a BCl$_{3}$/Cl$_{2}$ gas mixture, followed by a SF$_{6}$/C$_{4}$F$_{8}$ mixture to etch the pattern onto the silicon nitride film.

We resorted to the hard mask and two-stage etching because of the low selectivity between ZEP and silicon nitride of our RIE process that would not allow us to etch through the 369 nm thick membrane with just the ZEP mask. A final KOH wet etch is used to undercut the silicon wafer substrate, leaving the silicon nitride structure as a free standing film. The final mirror structure can be seen in Fig. 2a. The outer large trenches allow access to the underlying silicon wafer for the KOH solution to undercut the pattern, as well as to allow for easy removal of the mirror during the mounting procedure.
% by mostly detaching it from the surrounding membrane. 
The two dimensional lattice of PC holes is patterned as the square at the centre of the structure shown in Fig. 2a and is large enough to cover the hollow core and photonic crystal region of the fiber. 
%The larger outer holes surrounding the photonic-crystal region allow excess epoxy to escape when we mount the membrane onto the fiber tip in the procedure described below.

\section{Cavity assembly}

We mount a pair of membranes onto the ends of a HCPCF segment using vacuum compatible UV-curable epoxy (Norland Optical Adhesive 88). The mounting technique is based on the method developed by Shambat \textit{et. al} \cite{Shambat2011}. It is  summarized in Fig. 2b and consists of the following steps: (I.) A sharp ($\sim 50\mu m$ diameter) tungsten probe is first used to deposit two small epoxy droplets near the mirror template. This is achieved by dipping the tip of the probe into a drop of epoxy and then tapping and translating the probe on the surface of the chip until droplets of the desired size are produced. (II.) A pair of the droplets is then stamped onto the glass cladding area of the cleave face on one end of a HCPCF segment using a micromanipulator stage. (III.) The HCPCF is then aligned and lowered onto the PC mirror template and the setup is flooded with UV light to cure the epoxy. (IV.) The HCPCF is then retracted with the PC membrane attached. Depending on the intended application, the UV-curable epoxy can also be substituted by almost any two-component epoxy with a convenient curing time.
  
\begin{figure}[htb!]
\centering
%\begin{subfigure}[b]{0.75\linewidth}
%\includegraphics[width=\linewidth]{MountingScheme.pdf}
%\caption{}\label{fig:Mounting}
%\end{subfigure}
%\begin{subfigure}[b]{.43\linewidth}
%\includegraphics[width=\linewidth]{Template.pdf}
%\caption{}\label{fig:Template}
%\end{subfigure}
%\begin{subfigure}[b]{0.43\linewidth}
%\includegraphics[width=\linewidth]{Mounted_SEM.pdf}
%\caption{}\label{fig:MountedSEM}
%\end{subfigure}
\includegraphics[width=\linewidth]{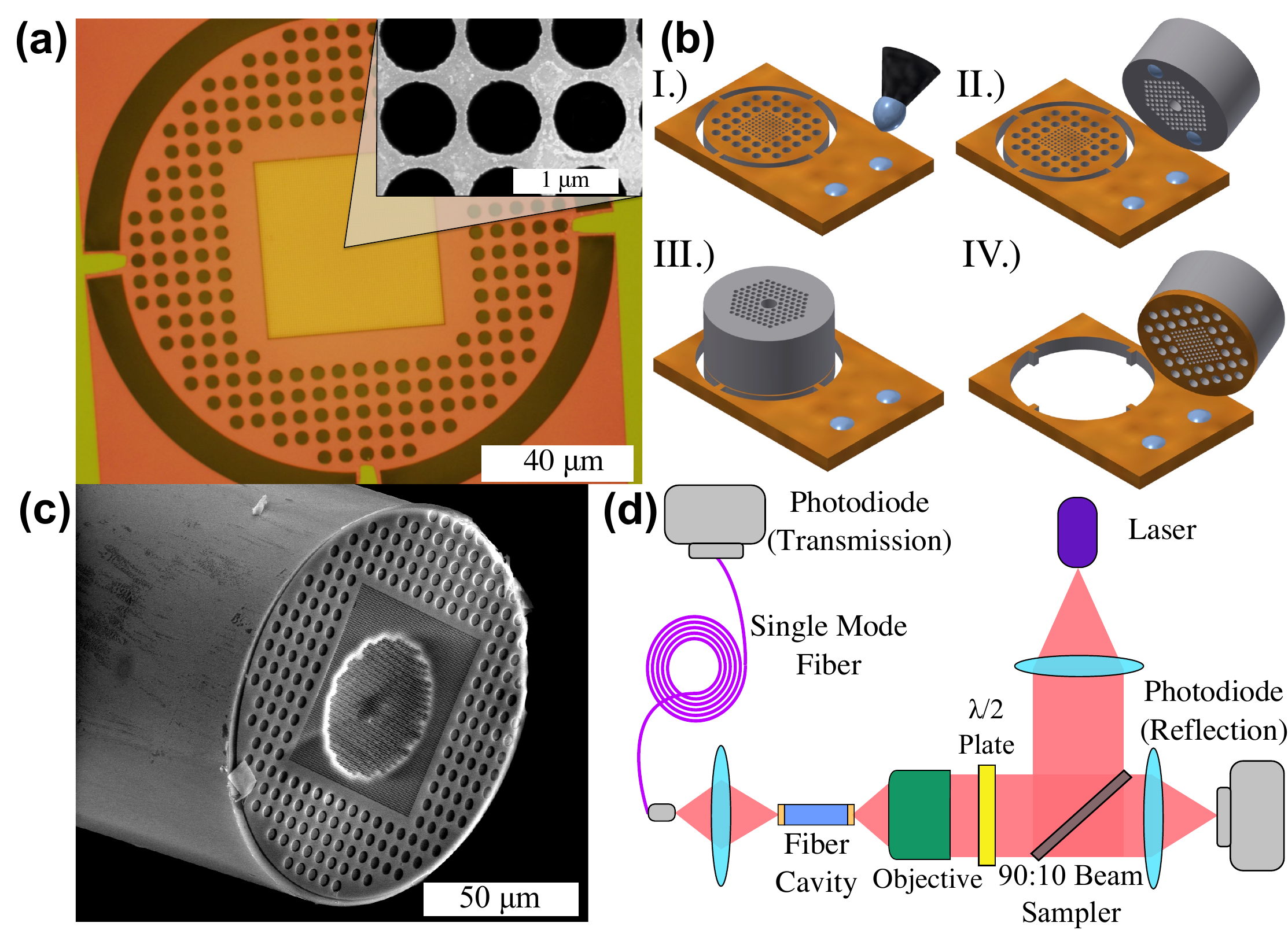}
\caption{(a) The fabricated mirror template with the PC pattern that will be mounted onto the HCPCF tips to form a cavity. The inset shows a scanning electron microscope (SEM) image of the inner square of photonic crystal holes at the desired dimensions to produce maximum reflectivity at $\sim$852 nm. (b) A diagram of the mounting technique. (c) SEM image of the PC membrane attached to the tip face of a HCPCF segment. The centre square lattice of PC holes in the membrane covers the entire PC region of the underlying HCPCF. (d) Optical setup used to measure the transmission and reflectivity spectrum of the fiber-cavity. 
%A single-mode fiber is used to filter out all higher order modes from the HCPCF and allow only the fundamental Gaussian mode to be detected. 
A half-wave ($\lambda/2$) plate is used to align the polarization of the input light with the slow and fast axis of the birefringent HCPCF.}
\end{figure}

The trenches that form the edge of the template allow for easy separation of the structure from the remaining silicon nitride film thanks to the narrow connecting bridges. The large outer holes that surround the inner square of the photonic-crystal region allow the epoxy to leak through the film and the membrane to sit flush against the fiber face. This also prevents excess epoxy from seeping into the PC region. Fig. 2c shows an SEM image of the PC membrane mounted onto the end of a HCPCF.

\section{Spectral Measurements of the Cavity}
The fiber cavity was mounted onto a piece of silicon wafer with a lithographically defined clamping structure \cite {Maruf2017} and its transmission spectra was measured using the optical setup shown in Fig. 2d. The light from a Ti:Sapph laser is aligned onto the fiber core using a 10x microscope objective and the transmitted light is coupled into a single mode fiber (SMF) before being detected by a photodiode. The SMF provides a spatial filter to remove any undesired higher order modes that may have been excited by the initial light coupling into the cavity so that the spectrum arising only from the fundamental mode of the fiber cavity can be observed.

\begin{figure}[htb!]
\centering
%\begin{subfigure}[b]{0.6\linewidth}
%\includegraphics[width=\linewidth]{Reflsetup.pdf}
%\caption{}\label{fig:Reflsetup}
%\end{subfigure}	
%\begin{subfigure}[b]{0.6\linewidth}
%\includegraphics[width=\linewidth]{fitspectrum.pdf}
%\caption{}\label{fig:fitspectrum}
%\end{subfigure}
%\begin{subfigure}[b]{.49\linewidth}
%\includegraphics[width=\linewidth]{Reflvswavelength.pdf}
%\caption{}\label{fig:Ref}
%\end{subfigure}
%\begin{subfigure}[b]{.49\linewidth}
%\includegraphics[width=\linewidth]{QfactorFinesse.pdf}
%\caption{}\label{fig:QF}
%\end{subfigure}
\includegraphics[width=\linewidth]{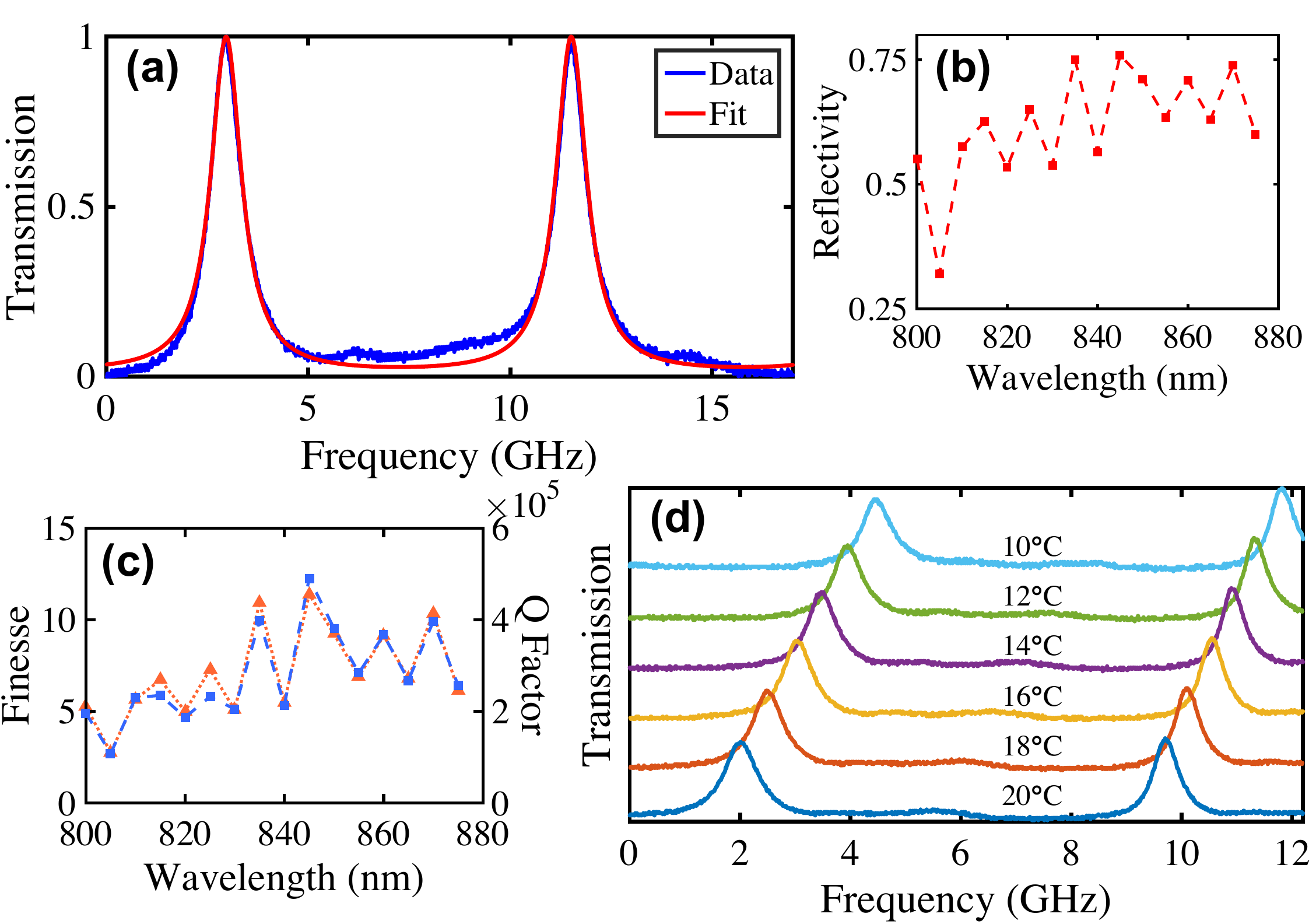}
\caption{(a) The transmission spectrum of the fiber-cavity found by scanning the input light frequency at a centre wavelength of 852nm. The normalized transmission data (blue line) is fitted to Eq. \ref{eq:transmission} (red line). (b) The reflectivity of the individual PC mirrors on the fiber tips extracted from the transmission spectrum fit. The parameters of the PC membrane measured from its SEM image were $a\approx$680 nm, $r\approx$260 nm, and $d\approx$363 nm. (c) Finesse (orange triangles) and Q factor (blue squares) of the fiber-cavity for a range of wavelengths. The finesse and FSR are found by fitting Eq. \ref{eq:transmission} to the cavity transmission spectrum and the Q factor is then found using the relationship $Q=\frac{\mathscr{F}}{FSR}\nu$. Connecting lines in (b) and (c) are for eye guidance. (d) Normalized transmission spectrum of the cavity at varying temperatures.}
\end{figure}

The cavity transmission spectrum, an example of which is shown in Fig. 3a, can be fitted to the normalized transmission intensity, $T$, of a Fabry-P\'{e}rot resonator:

\begin{equation}
T=\frac{I}{I_{0}}=\frac{1}{1+\left(\frac{2\mathscr{F}}{\pi}\right)^{2}\sin^{2}\left(\frac{\pi\nu}{FSR}\right)},
\label{eq:transmission}
\end{equation}
where $\nu$ is the frequency if the incident light. The two fitting parameters that are determined from this spectrum are the finesse, $\mathscr{F}$, and the free spectral range, $FSR=\frac{c}{2nL}$, where $c$ is the speed of light, $L$ is the length of the cavity, and $n$ is the effective refractive index of the cavity medium. The finesse and the FSR then determine the quality factor of the resonator, $Q=\frac{\mathscr{F}}{FSR}\nu$.

In addition to quantifying the performance of our cavity, measuring the finesse also allows us to evaluate the power reflectivity, $R$, of the photonic-crystal mirrors. The finesse is determined by the round trip loss in the electric field, $r$, such that
\begin{equation}
\mathscr{F}=\frac{\pi\sqrt{r}}{1-r}
\label{eq:finesse}
\end{equation}
in which each round trip field loss is $r=\sqrt{R_1}\sqrt{R_2}e^{-\frac{\alpha d}{2}}$, for mirror power reflectivities, $R_1$ and $R_2$, round trip distance, $d$, and propagation loss, $\alpha$. Assuming identical mirrors, $R_1=R_2\equiv R$, and a round trip distance of twice the cavity length, the round trip loss becomes
\begin{equation}
r=Re^{-\alpha L}
\label{eq:r}
\end{equation}
For the case of the HC-800-02 fiber used to assemble the cavity, this loss is $\sim 150$dB/km for wavelengths around 850nm. Using the fitted values for $FSR$ and $\mathscr{F}$ from the cavity transmission spectrum, the reflectivities of the cavity PC mirrors can then calculated using Eq. \ref{eq:finesse} and \ref{eq:r}, as shown in Fig. 3b, giving a maximum $R\sim76\%$ at 845nm. 
%We are currently trying to optimize our fabrication process to bring this value into agreement with the reflectivity predicted by our numerical simulations  ($\sim99\%$).

It should be mentioned that the fiber-cavity can actually produce two separate peaks in the transmission spectrum. This is due to the fact that the HCPCF is slightly birefringent and thus different polarizations will experience different propagation constants in the fiber. This results in slightly different $FSR$s for polarizations aligned with the fast and the slow axis of the fiber, which produces two transmission peaks when the polarization of the input light is not aligned along one of the birefringence axes. This issue can be resolved by simply rotating the polarization of the incident light coupling into the fiber-cavity with a half-wave plate
%as depicted in Fig. \ref{fig:Reflsetup}, until
% the polarization is completely along either the fast or slow axis and thus
until only one polarization mode of the cavity is excited.

%To evaluate the power reflectivity, $R$, of each of the PC mirrors, the absorption loss due to the fiber must be accounted for in order to determine the true reflectivity. This is done by using the expression for the Q factor of a Fabry-P\'{e}rot cavity, given by Eq. \ref{eq:qfactcavity}.

%In order to characterize the performance of our cavities, we determine the finesse (from fitting the cavity transmission spectrum) and the Q factor (from Eq. \ref{eq:qfact}), which essentially describes the rate at which energy is dissipated from the cavity. Fig. \ref{fig:QF} shows both the finesse (orange dotted line) and Q factor (blue dashed line) at varying wavelengths, with a maximum value of $\sim11$ and $\sim4.9\times10^{5}$ at 845nm, respectively.

%\begin{figure}[htb!]
%\centering
%\includegraphics[width=.6\linewidth]{CavityTuning.pdf}
%\caption{Normalized transmission spectrum of the cavity at varying temperatures. 
%%from $10^\circ$C to $20^\circ$C.
%}
%\label{fig:tuning}
%\end{figure}

The reflectivities of our fabricated PC membranes are for now significantly below the design prediction.
%, which can be mostly attributed to the dimensions of the PC holes not matching our simulated design specifications due to the limits in our fabrication process. 
However, refinements to our fabrication procedure should allow us to produce mirrors with much higher reflectivities and, consequently, cavities with increased finesse, as reflectivities exceeding 99$\%$ have been demonstrated \cite{Chen2017}, albeit for mirrors designed for beams with larger waist of $\sim$ 50$\mu$m. 
%The currently measured finesse and Q factors of our fiber-cavity will also, of course, increase as a result of utilizing PC mirrors with larger reflectivity values. 
The currently fabricated mirrors form fiber cavities with maximum $\mathscr{F}\approx12$ and Q-factor of about 5$\times$10$^5$ (Fig. 3c).

Finally, Fig. 3d shows tuning of our cavity by changing its temperature. For this measurement, the silicon wafer piece holding the fiber cavity was mounted onto a thermoelectric element.
%The spectral transmission peaks of our fiber-cavity was also shown to be tunable by altering the temperature of the cavity. 
As the cavity temperature is changed, the thermal expansion changes the length of the HCPCF, $L$, by about 5-10 nm per $^\circ$C, causing the cavity resonances to shift by $\sim0.2$ GHz per $^\circ$C at the measured optical frequency ranges.
 In practical applications, the cavity can thus either be stabilized by keeping it at a specific temperature or locked to a reference frequency using temperature tuning as part of the feedback loop.

\section{Conclusion and outlooks}

In conclusion, we have demonstrated a novel type of fiber-integrated cavity created by attaching a pair of perforated photonic membranes acting as dielectric metasurface mirrors to the tips of a HCPCF piece. As the holes can allow injection of gases into the fiber, we expect this cavity to open paths to a broad range of exciting new applications. Even with the currently demonstrated relatively low finesse of $\sim$10, the cavity can be used to improve the performance of fiber-integrated gas lasers \cite{HOFGLAS2012} and frequency references \cite{Wang2013}. While not yet confirmed through numerical simulations, the membrane should in principle also allow the loading of laser-cooled atoms into the HCPCF. The tight confinement of photons over macroscopic distances in the HCPCF cavity, not available in confocal cavities, could potentially enable novel regimes of the recently demonstrated super-radiant lasers \cite{Bohnet2012} and spin-squeezing measurements \cite{Leroux2010, Hosten2016}, as well as offer an alternative to tapered fibers in exploring multi-mode strong-coupling cavity quantum electrodynamics (cQED) systems \cite{Schneeweiss2017}. This fiber-based cavity also allows for the potential of on-chip integration by the use of lithographically fabricated mechanical structures forming fiber couplers \cite{Maruf2017}.

With improved reflectivities of the fabricated membranes, a cavity with high single-photon cooperativity \cite{Tanji-review2011} could be realized in the very near future \cite{Bappi2017} and pave the way toward tantalizing new studies combining coherent control techniques of light propagation, such as slow light and light storage, with cQED phenomena in atomic ensembles. In particular, these perforated membranes can be designed as polarization selective mirrors \cite{Lousse2004, Chang-review2012} that reflect light with, e.g., vertical polarization but transmit light polarized horizontally and such mirrors can then form polarization-selective cavities inside the HCPCF. In turn, a polarization selective cavity can be used to implement vacuum-induced transparency (VIT) \cite{Tanji2011} in a high-optical-depth ensemble or to improve the performance of the recently demonstrated all-optical transistor controlled by a single photon \cite{Chen2013} by simplifying its geometry and allowing the 'gate' and 'source' fields to both propagate along the axis of the cavity.    

\begin{acknowledgement}

The authors thank Industry Canada and by Canada's Natural Sciences and Research Council (NSERC) under the Discovery Grants Program. J. F. has been in part supported by Ontario Graduate Scholarship (OGS). The authors also thank the Quantum Nanofab facility for their support with the nanofabrication part of this work, in particular Nathan Nelson-Fitzpatrick for growing the LPCVD SiN films.

\end{acknowledgement}


\bibliography{PCFcavity-membrane}

\end{document}